\documentstyle[12pt,epsf,fleqn]{article}
\newcommand{\Integer}{\:\mbox{\sf Z} \hspace{-0.82em} \mbox{\sf Z}\,}

\def\nsection#1{\setcounter{equation}{0}\section{#1}}

\def\e{\mbox{e}}
\def\ib{\,\mbox{i}\,}
\def\is{\,\mbox{\scriptsize i}\,}
\def\la{\lambda}
\def\th{\hbox{th}}
\def\te{\vartheta_1}

\def\tv{\vartheta_4}
\def\case#1#2{{\textstyle{#1\over #2}}}
\def\W#1#2#3#4#5{W #1 \! \left(\hspace{-1mm}
         \begin{array}{cc}#5 & #4 \\ #2 & #3 \end{array}
         \hspace{-1mm}\right)}
\def\case#1#2{{\textstyle{#1\over #2}}}

\begin{document}

\title{Lattice Ising model in a field: E$_8$ scattering theory}

\author{
V. V. Bazhanov\thanks{
IAS, Australian National University,
Theoretical Physics and Mathematics,
GPO Box 4, Canberra, ACT 2601, Australia, e-mail {\tt vvb105@phys.anu.edu.au}}
\thanks{
On leave of absence from the Institute
for High Energy Physics,
Protvino, Moscow Region, 142284, Russia.} ,
B. Nienhuis\thanks{
Instituut voor Theoretische Fysica,
Universiteit van Amsterdam,
Valckenierstraat~65,
1018~XE Amsterdam,
The Netherlands, e-mail {\tt nienhuis@phys.uva.nl}}
\hspace{0.2mm} and
S. O. Warnaar$^{\mbox{\footnotesize\ddag}}$\thanks{
Present address: Mathematics Department,
University of Melbourne, Parkville, Victoria 3052,
Australia, e-mail {\tt warnaar@mundoe.maths.mu.oz.au}}}

\date{\ }
\maketitle

\begin{abstract}
Zamolodchikov found an integrable field theory
related to the Lie algebra E$_8$, which describes the
scaling limit of the Ising model in a magnetic field.
He conjectured that there also exist solvable lattice
models based on E$_8$ in the universality class of the Ising model in a
field.
The dilute A$_3$ model is a solvable lattice model with
a critical point in the Ising universality class.
The parameter by which the model can be taken away
from the critical point acts like a magnetic field by breaking
the $\Integer_2$ symmetry between the states.
The expected direct relation of the model with E$_8$
has not been found hitherto.
In this letter we study the thermodynamics of the dilute
A$_3$ model and show that in  the scaling limit it exhibits
an appropriate E$_8$ structure, which naturally leads to the E$_8$
scattering theory for massive excitations over the ground state.
\end{abstract}

\newlength{\mathin}
\setlength{\mathin}{\mathindent}
\setcounter{page}{1}
\nsection{Introduction}
Since the work \cite{Zamolodchikov-A87b}
by A.B. Zamolodchikov it is known that certain
perturbations of conformal field theories (CFT's) lead to completely
integrable models of massive quantum field theory (QFT).
The
existence of non-trivial higher integrals
of motion and other dynamical symmetries
\cite{Zamolodchikov-A88a,A-B-LC,Reshetikhin90d,LeClair,Eguchi}
in such a QFT allows to compute the spectrum of the particles
and their $S$-matrix explicitly.
At the same time, these QFT models can be obtained as the
scaling limit of appropriate non-critical solvable lattice models
in statistical mechanics
(see \cite{Baxter-book} for an introduction
and references on solvable lattice models).
In the latter  approach the spectrum and  the $S$-matrices
can be calculated from the Bethe Ansatz equations for the
corresponding lattice model
\cite{Reshetikhin87a,Bazhanov90d,Reshetikhin91c}.
The natural problem arising in this connection is to find lattice
counterparts for all known integrable perturbed CFT's and
vice versa. A description of known results of such
correspondence lies outside the scope of this letter and we refer the
interested reader to [1-10]
and references therein.
Here we consider one
particularly important example of this correspondence associated with
the Ising model at its critical temperature in a magnetic field,
hereafter referred to as the magnetic Ising model.

A.B. Zamolodchikov has shown \cite{Zamolodchikov-A89a}
that the $c=1/2$ CFT (corresponding to the critical
Ising  model) perturbed with the spin operator $\phi_{1,2}=\phi_{2,2}$
of dimension $(1/16,1/16)$ describes an exactly integrable QFT
containing eight massive particles with a reflectionless factorised
$S$-matrix. Up to normalisation the masses of these particles coincide
with the components $S_i$ of the Perron-Frobenius vector of the Cartan
matrix of the Lie algebra E$_8$:
\begin{equation}
\frac{m_i}{m_j} = \frac{S_i}{S_j}.
\label{massratio}
\end{equation}
The element of the
$S$-matrix describing the scattering of the lightest particles, with
mass $m_1$,
reads \cite{Zamolodchikov-A89a}
\begin{equation}
S_{1,1}(\beta)=\frac
{\th\left(\frac{\beta}{2} + \ib \frac{\pi}{6} \right)
 \th\left(\frac{\beta}{2} + \ib \frac{\pi}{5} \right)
 \th\left(\frac{\beta}{2} + \ib \frac{\pi}{30} \right)}
{\th\left(\frac{\beta}{2} - \ib \frac{\pi}{6} \right)
 \th\left(\frac{\beta}{2} - \ib \frac{\pi}{5} \right)
 \th\left(\frac{\beta}{2} - \ib \frac{\pi}{30} \right)},
\end{equation}
with $\beta$ the rapidity.
The other elements are uniquely determined by the
bootstrap program \cite{Zamolodchikov-A89a}.

The aim of this letter is to show
that the above QFT describes the scaling limit of
the dilute A$_3$ model of
Warnaar, Nienhuis and Seaton \cite{WNS,WPSN}
in the appropriate regime.
It should be noted that there were some earlier, rather strong
indications supporting the above correspondence.
All these parts remarkably fit together with our results,
completing a sequence of arguments which can be summarised as follows:
\begin{description}
\item[{\rm (i)}]
The dilute A$_3$ model is an interaction-round-a-face model on
the square lattice with spins taking three values (detailed
definitions are given in equations
(\ref{Incidence})-(\ref{regimes})). Admissible values of the
adjacent spins  are
determined by the incidence matrix (\ref{Incidence}),
which has largest eigenvalue
equal to $1+\sqrt{2}$.
\item[{\rm (ii)}]
The model has two
physically distinct regimes of relevance to our discussion,
here denoted as {\it i)} and {\it ii)},
depending on the region of the spectral parameter or, equivalently, of
a sign of the Hamiltonian of the associated one-dimensional chain.
(These are the regimes $2^+$ and $3^+$ of ref~\cite{WPSN}, respectively).
The central charges and the conformal dimensions of the leading perturbation
computed from exact expressions for the free energy and the local
state  probabilities
of the dilute A$_3$ model
for these two regimes read \cite{WBN,WPSN}
\begin{equation}
i) \quad c=1/2, \quad \Delta=1/16; \qquad ii) \quad c=6/5, \quad
\Delta=15/16.
\label{canddelta}
\end{equation}
\item[{\rm (iii)}]
In ref~\cite{Bazhanov90d,Bazhanov90b} Bazhanov and Reshetikhin
proposed thermodynamic Bethe Ansatz equations (TBAE) related
to the A-D-E Lie algebras, corresponding to
non-critical models in statistical mechanics.
Using standard
thermodynamics calculations and the high level Bethe Ansatz (see
\cite{Reshetikhin87a} and references therein)
they computed: the central charges of the corresponding scaling field
theories, dimensions of the leading
perturbations, the spectra and scattering amplitudes of the
massive excitations, expressing them through fused Boltzmann weights.
In particular, in the  case relevant to  our discussion
($\cal G$=E$_8$, $g=30$, $p=\ell=1$,
in the notation of \cite{Bazhanov90d}) the
exponents they found \footnote{Note that the equations (5$\cdot$1)
and (5$\cdot$4) in \cite{Bazhanov90d}
have been misprinted. Correcting (5$\cdot$1)
to $c=c^{\cal G}(l)+c^{\cal G}(r-l-g)-c^{\cal G}(r-g)+
\mbox{rank}\:{\cal G}$ yields the following result for
the central charge in (5$\cdot$4):
$c=2\:\mbox{rank}\:{\cal G}/(g+2)$. Also the phrases
``minimal unitary'', just before, and
``by the operator $\phi_{(1,3)}$'' just after (5$\cdot$4)
should be deleted.}
precisely match (\ref{canddelta})
in both regimes.
Furthermore, the TBAE   allowing the calculation of
the largest eigenvalue of the incidence matrix of the underlying
lattice model, gave in this case precisely the value $1+\sqrt{2}$
\cite{Bazhanov-remark-in-Kuniba's-paper}.
\item[{\rm (iv)}]
Finally, the
spectrum  and $S$-matrix of the scaling field theory in regime
$\it i)\/$ found in \cite{Bazhanov90d} from the
high level Bethe Ansatz for E$_8$
coincide with those of Zamolodchikov's magnetic Ising model.
\end{description}
All the above arguments strongly suggest that the TBAE
based on the Lie algebra E$_8$ as
proposed in \cite{Bazhanov90d,Bazhanov90b},
are those of the the dilute A$_3$ model.

In this paper we present the Bethe Ansatz equations (BAE)
for the non-critical, dilute A$_L$ model.
As these equations, at criticality, are very similar to those of the
Izergin-Korepin model \cite{IK,Vichirko}, it is not
surprising that, when specialised to $L=3$,
they do not display any explicit structure related
to the root system of E$_8$.
It turns out however that this structure reveals itself
in a quite  complicated string structure of the solutions
to the BAE.
Motivated by an extensive numerical
investigation of the BAE
we formulate an exact conjecture
for the thermodynamically significant strings.
This leads to TBAE, which, rewritten in a
new string basis precisely yield the E$_8$ based TBAE of
ref~\cite{Bazhanov90d} discussed under (iii).
As a result of (iv) this
finalises the
correspondence between the dilute A$_3$ model and the
magnetic Ising model.

\nsection{The dilute A models}
The dilute A$_L$ model, belonging to the more
general class of dilute A-D-E models,
is an exactly solvable, restricted solid-on-solid
model defined on the square lattice.
Each site of the lattice can take one of $L$ possible
(height) values, subject to the restriction that
neighbouring sites of the lattice either have the
same height, or differ by $\pm 1$.
This adjacency condition can be
conveniently expressed by a so-called
incidence matrix $M$:
\begin{equation}
M_{a,b} = \delta_{a,b-1} + \delta_{a,b} + \delta_{a,b+1}
\qquad a,b\in \{1,\ldots,L\},
\label{Incidence}
\end{equation}
where we note that $M$ relates to the
Cartan matrix $C^{\mbox{\scriptsize A}_L}$ of
the Lie algebra A$_L$
by $M=3 I - C^{\mbox{\scriptsize A}_L}$,
with $I$ the identity matrix.
The eigenvalues of the incidence matrix are
found to be
\begin{equation}
\Lambda_j = 1 + 2\cos
\left(\frac{\pi j}{L+1} \right) \qquad j=1,\ldots,L.
\end{equation}
For the case of interest here, $L=3$, we thus find the
largest eigenvalue to be $1+\sqrt{2}$, in accordance with
the prediction for the E$_8$ TBAE as mentioned in (iii)
of the introduction.

Using standard definitions of $\vartheta_{i}(u,q)$-functions,
suppressing the dependence on the nome $q=\e^{-\tau}$, $\tau>0$,
the Boltzmann weights of the allowed height configurations of
an elementary face of the lattice are
\setlength{\mathindent}{0 cm}
\begin{eqnarray}
\lefteqn{\W{}{a}{a}{a}{a}=
\frac{\te(6\la-u)\te(3\la+u)}{\te(6\la)\te(3\la)}}
\nonumber \\  & & \nonumber \\
\lefteqn{\hphantom{\W{}{a}{a}{a}{a}}
-\left(\frac{S(a+1)}{S(a)}\frac{\tv(2a\la-5\la)}{\tv(2a\la+\la)}
      +\frac{S(a-1)}{S(a)}\frac{\tv(2a\la+5\la)}{\tv(2a\la-\la)}\right)
\frac{\te(u)\te(3\la-u)}{\te(6\la)\te(3\la)}}
\nonumber \\ & & \nonumber \\
\lefteqn{\W{}{a}{a}{a}{a\pm 1}=\W{}{a}{a\pm 1}{a}{a}=
\frac{\te(3\la-u)\tv(\pm 2a\la+\la-u)}{\te(3\la)\tv(\pm 2a\la+\la)}}
\nonumber \\ & & \nonumber \\
\lefteqn{\W{}{a\pm 1}{a}{a}{a}=\W{}{a}{a}{a\pm 1}{a}=
\left(\frac{S(a\pm 1)}{S(a)}\right)^{1/2}
\frac{\te(u)\tv(\pm 2a\la-2\la+u)}{\te(3\la)\tv(\pm 2a\la+\la)}}
\nonumber \\ & & \nonumber \\
\lefteqn{\W{}{a}{a\pm 1}{a\pm 1}{a}=\W{}{a}{a}{a\pm 1}{a\pm 1}}
\nonumber \\ & & \nonumber \\
\lefteqn{ \hphantom{\W{}{a}{a\pm 1}{a\pm 1}{a}}
=\left(\frac{\tv(\pm 2a\la+3\la)\tv(\pm 2a\la-\la)}
           {\tv^2(\pm 2a\la+\la)}\right)^{1/2}
\frac{\te(u)\te(3\la-u)}{\te(2\la)\te(3\la)} }
\nonumber \\ & & \label{Bweights} \\
\lefteqn{\W{}{a}{a\mp 1}{a}{a\pm 1}=
\frac{\te(2\la-u)\te(3\la-u)}{\te(2\la)\te(3\la)}}
\nonumber \\ & & \nonumber \\
\lefteqn{\W{}{a\pm 1}{a}{a\mp 1}{a}=
-\left(\frac{S(a-1)S(a+1)}{S^2(a)}\right)^{1/2}
\frac{\te(u)\te(\la-u)}{\te(2\la)\te(3\la)}}
\nonumber \\ & & \nonumber \\
\lefteqn{\W{}{a\pm 1}{a}{a\pm 1}{a}=
\frac{\te(3\la-u)\te(\pm 4a\la+2\la+u)}{\te(3\la)\te(\pm 4a\la+2\la)}
+\frac{S(a\pm 1)}{S(a)}
\frac{\te(u)\te(\pm 4a\la-\la+u)}{\te(3\la) \te(\pm 4a\la+2\la)}}
\nonumber \\ & & \nonumber \\
\lefteqn{\hphantom{\W{}{a\pm 1}{a}{a\pm 1}{a}}=
\frac{\te(3\la+u)\te(\pm 4a\la-4\la+u)}
{\te(3\la)\te(\pm 4a\la-4\la)}}
\nonumber \\ & & \nonumber \\
\lefteqn{\hphantom{\W{}{a\pm 1}{a}{a\pm 1}{a}}+
\left(\frac{S(a\mp 1)}{S(a)}\frac{\te(4\la)}{\te(2\la)}
-\frac{\tv(\pm 2a\la-5\la)}{\tv(\pm 2a\la+\la)} \right)
\frac{\te(u)\te(\pm 4a\la-\la+u)}{\te(3\la) \te(\pm 4a\la-4\la)}}
\nonumber \\ & & \nonumber \\
\lefteqn{S(a)=(-)^{\displaystyle a} \;
\frac{\te(4a\la)}{\tv(2a\la)} \, .}
\nonumber
\end{eqnarray}
The variable $\lambda$ and the range of the spectral parameter $u$
in the above weights are
given by\footnote{In \cite{WPSN} two more regimes were
defined, which are omitted being of no relevance here.}
\setlength{\mathindent}{\mathin}
\begin{equation}
\lambda = \frac{\pi}{4} \, \frac{L+2}{L+1}
\qquad
\left\{
\begin{array}{lll}
0<u<3\lambda   & \qquad & \mbox{regime } i) \\
3\lambda-\pi<u<0   & \qquad & \mbox{regime } ii) .
\end{array}
\right.
\label{regimes}
\end{equation}

\nsection{Bethe Ansatz}
The transfer matrix of the dilute A models is
defined in the usual way as
\begin{equation}
T_{\{a\}}^{\{b\}} = \prod_{j=1}^{N}
\W{}{a_j}{a_{j+1}}{b_{j+1}}{b_j} ,
\end{equation}
where $\{a\}$ is an admissible path of heights and
$a_{N+1} =a_1$, $b_{N+1} = b_1$.
The number of admissible paths is given by Trace$(A^N)=
\sum_{j=1}^L \Lambda_j^N$.

We define two positive integers $p$ and $r$ and a renormalised
spectral parameter $\theta$ by
\begin{equation}
\lambda =   \frac{p}{r} \; \pi \qquad
u = \frac{\theta}{2 r} \; \pi     ,
\end{equation}
where $p$ and $r$ are coprime.
Furthermore we introduce the modified $\vartheta$-function
\begin{equation}
h(\theta) = \frac{1}{2} \; q^{-1/4} \;
\te\left(\frac{\pi \theta}{2 r},q \right).
\end{equation}
With these definitions the eigenvalues of the transfer matrix are
found to be
\begin{eqnarray}
\Lambda(\theta) &=& \omega \left(
\frac{h(4p-\theta)\;h(6p-\theta)}{h(4p)\; h(6p)}\right)^N
\prod_{j=1}^N
\frac{h(\theta+\ib\theta_j+2 p)}{h(\theta+\ib\theta_j-2 p)}
\nonumber \\
&+& \left(
\frac{h(\theta)\;h(6p-\theta)}{h(4p)\; h(6p)}\right)^N
\prod_{j=1}^N
\frac{h(\theta+\ib\theta_j) \; h(\theta+\ib\theta_j-6 p)}
     {h(\theta+\ib\theta_j-2 p) \; h(\theta+\ib\theta_j-4 p)} ,
\label{Eival}
\\
&+& \omega^{-1}
\left(
-\frac{h(\theta)\;h(2p-\theta)}{h(4p)\; h(6p)}\right)^N
\prod_{j=1}^N
\frac{h(\theta+\ib\theta_j-8 p)}
     {h(\theta+\ib\theta_j-4 p)} \nonumber
\end{eqnarray}
where the numbers $\{\theta_j\}$ are given by the following
set of BAE:
\setlength{\mathindent}{0 cm}
\begin{equation}
\omega \left(
\frac{h(2 p + \ib \theta_j)}
     {h(2 p - \ib \theta_j)}\right)^{N} =
-\prod_{k=1}^{N}
\frac{h(\ib \theta_j - \ib \theta_k + 4 p) \;
      h(\ib \theta_j - \ib \theta_k - 2 p)}
     {h(\ib \theta_j - \ib \theta_k - 4 p) \;
      h(\ib \theta_j - \ib \theta_k + 2 p)} \quad j=1,\ldots,N
\label{BAE}
\end{equation}
and $\omega=\exp(\ib \pi \ell/(L+1))$, $\ell=1,\ldots,L$.
Note that if we set the nome $q$ of the function
$h$ equal to zero and choose $\omega$ to be unity,
the above eigenvalue expression and BAE
reduce to those of the
Izergin-Korepin model \cite{Vichirko} in the sector in which the
number of roots is equal to the system size $N$.
A proof of equations (\ref{Eival}) and (\ref{BAE}) in the
critical limit, based
on an extension of the mapping of RSOS models onto vertex models
found in \cite{WN}, will be presented elsewhere \cite{BNW}.
In the general elliptic case the equations can be proven
using the functional equations for the dilute A models
as obtained in ref~\cite{GPZ}.

\nsection{Thermodynamic Bethe Ansatz}
The remaining part of this letter is devoted to
the solution of the
BAE in the thermodynamic limit for the dilute A$_3$ model,
corresponding to the case $p=5$ and $r=16$.

Let $n^{(t)}$ be a positive integer,
$\Delta^{(t)}$ an
$n^{(t)}$-dimensional
vector with integer coefficients $\Delta_k^{(t)}$
and $\varepsilon^{(t)}=0,1$.
We then define a string of type $t$ as a set of
complex numbers
$\{\alpha_{j,k}^{(t)}\}$ with
\setlength{\mathindent}{\mathin}
\begin{equation}
\alpha_{j,k}^{(t)} = \alpha_j^{(t)} +
\ib (\Delta_k^{(t)} + \varepsilon^{(t)} r )
\qquad k=1,\ldots,n^{(t)} ,
\end{equation}
where the real number $\alpha_j^{(t)}$ specifies the centre
of the string.
Based on an extensive numerical study \cite{BNW} we find that
for $N\to\infty$ each
solution $\{\theta_j\}$ of
(\ref{BAE}) consists of a collection of strings, where we have {\it only
nine thermodynamically significant string types} (in the sense that
$N^{(t)}/N$ is finite for $N\to\infty$)
\begin{equation}
N = \sum_{t=0}^8  n^{(t)} N^{(t)}+o(N) ,
\label{Nsum}
\end{equation}
with $N^{(t)}$ the total number of strings of type $t$.
These nine allowed string types are listed in
Table~1. We expect the above
statement to be exact and claim it as a conjecture.
\begin{table}[hbt]
\centering
$$
\begin{array}{|c|c|c|c|}
\hline
t & n^{(t)} & \Delta^{(t)}/5 & \varepsilon^{(t)} \\
\hline
0 & 1  & (0) & 0 \\
1 & 2  & (-1,1) & 1 \\
2 & 4  & (-4,-2,2,4) & 0 \\
3 & 6  & (-7,-5,-1,1,5,7) & 1 \\
4 & 8  & (-10,-8,-4,-2,2,4,8,10) & 0 \\
5 & 10 & (-13,-11,-7,-5,-1,1,5,7,11,13) & 1 \\
6 & 7  & (-14,-6,-2,0,2,6,14) & 1 \\
7 & 4  & (-3,-1,1,3) & 0 \\
8 & 5  & (-12,-8,0,8,12) & 1 \\
\hline
\end{array}
$$
\caption{The nine thermodynamically significant string types}
\label{Table}
\end{table}

In the thermodynamic limit the centers of the strings form continuous
distributions and the BAE (\ref{BAE}) lead to integral equations
for the densities of strings $\rho_t(\alpha)$ and ``holes''
$\tilde{\rho}_t(\alpha)$ \cite{Yang-Yang}
\begin{equation}
b_t(\alpha) = -(-)^{\delta_{t,0}}\tilde{\rho}_t(\alpha)
+\sum_{s=0}^8 B_{t,s} \ast \rho_s \, (\alpha)
\qquad t=0,\ldots,8,
\label{TBAE}
\end{equation}
where $a\ast b$ denotes the convolution of
the functions $a$ and $b$
\begin{equation}
a\ast b \, (\alpha) = \int_{-\tau r/\pi}^{\tau r/\pi}
a(\alpha-\beta) \, b(\beta) \: d\beta .
\end{equation}
The functions
$b_t$ and $B_{t,s}$ in (\ref{TBAE}),
which are $2\tau r/\pi$-periodic, read
\begin{eqnarray}
b_t(\alpha) &=& \sum_{k=1}^{n^{(t)}}
\psi_{2p} \left(\alpha + \ib \Delta_k^{(t)} \right)
\nonumber \\
B_{t,s}(\alpha) &=& -(-)^{\delta_{t,0}}
\delta_{t,s} \delta(\alpha)
+\sum_{k=1}^{n^{(t)}} \sum_{\ell=1}^{n^{(s)}}
\left[
\psi_{4p}\left(\alpha+\ib \Delta_k^{(t)} + \ib \Delta_l^{(s)} \right)
\right. \\
& & \qquad \qquad \qquad \qquad \qquad \qquad \left.
\vphantom{\sum_{k=1}^{n^{(t)}}}
-\psi_{2p}\left(\alpha+\ib \Delta_k^{(t)} + \ib \Delta_l^{(s)} \right)
\right], \nonumber
\end{eqnarray}
with $\psi_k$ defined as
\begin{equation}
\psi_k(\alpha) = \frac{1}{2\pi \ib} \frac{\mbox{d}}{\mbox{d}\alpha}
\log \left[
\frac{h(k+\ib\alpha)}{h(k-\ib\alpha)}\right].
\end{equation}
The function $\psi_k$ has the following Fourier
transform (FT)
\begin{equation}
\hat{\psi}_k(x) =  \frac{\sinh(r-k)x}{\sinh rx}
\qquad 0< k < 2r,
\end{equation}
where
\begin{eqnarray}
\hat{F}(x) &=& \int_{-\tau r/\pi}^{\tau r/\pi}
\e^{-\is \alpha x} F(\alpha) \: d\alpha \nonumber \\
F(\alpha) &=& \frac{\delta}{2\pi}
\sum_{n=-\infty}^{\infty} \e^{\is \alpha x_n} \hat{F}(x_n) ,
\end{eqnarray}
with $\delta=\pi^2/(r\tau)$, $x_n=\delta n$.

As usual, we define the (local) Hamiltonian $\cal H$
of the associated one-dimensional
integrable spin chain as the logarithmic derivative of
the transfer matrix at $\theta=0$.
Then, after appropriate normalisation and shift, the spectrum of $\cal H$,
in the limit $N\to\infty$ reads
\begin{equation}
\frac{E}{N} = \epsilon \sum_{t=0}^{8}
\int_{-\tau r/\pi}^{\tau r/\pi}
b_t(\alpha) \, \rho_t(\alpha) \: d\alpha +
O\left(\e^{-\mu N} \right) \quad \mu >0.
\label{Energy}
\end{equation}
where $\epsilon=-1$ for regime {\it i)} and
$\epsilon=1$ for regime {\it ii)} in (\ref{regimes}).

The densities $\rho_t$ are normalised such
that
\begin{equation}
\int_{-\tau r/\pi}^{\tau r/\pi}
\rho_t(\alpha)\: d\alpha = N^{(t)}/N.
\end{equation}
Therefore, from equation (\ref{Nsum}), we have
\begin{equation}
\sum_{t=0}^8
\int_{-\tau r/\pi}^{\tau r/\pi}
n^{(t)} \rho_t(\alpha) \: d\alpha = 1.
\end{equation}
This relation together with equation (\ref{TBAE}) for $t=0$
implies
\begin{equation}
\tilde{\rho}_0(\alpha) =0.
\label{zero}
\end{equation}
Hence we conclude that the strings of type 0 have no holes in any
state, and we eliminate $\rho_0(\alpha)$ from (\ref{TBAE}).
After a tedious calculation we find that the resulting integral
equations can naturally be described in
terms of the E$_8$ root system as follows.

Let $C^{\mbox{\scriptsize E}_8}_{t,s}$ $t,s=1,\ldots,8$
be the elements of the Cartan matrix for
E$_8$, where we use the following enumeration of the nodes of the
corresponding Dynkin diagram:
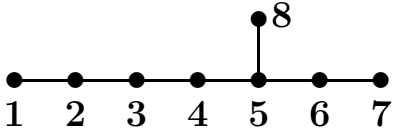
\[
\setlength{\unitlength}{0.008in}%
\begingroup\makeatletter
\def\x#1#2#3#4#5#6#7\relax{\def\x{#1#2#3#4#5#6}}%
\expandafter\x\fmtname xxxxxx\relax \def\y{splain}%
\ifx\x\y   
\gdef\SetFigFont#1#2#3{%
  \ifnum #1<17\tiny\else \ifnum #1<20\small\else
  \ifnum #1<24\normalsize\else \ifnum #1<29\large\else
  \ifnum #1<34\Large\else \ifnum #1<41\LARGE\else
     \huge\fi\fi\fi\fi\fi\fi
  \csname #3\endcsname}%
\else
\gdef\SetFigFont#1#2#3{\begingroup
  \count@#1\relax \ifnum 25<\count@\count@25\fi
  \def\x{\endgroup\@setsize\SetFigFont{#2pt}}%
  \expandafter\x
    \csname \romannumeral\the\count@ pt\expandafter\endcsname
    \csname @\romannumeral\the\count@ pt\endcsname
  \csname #3\endcsname}%
\fi
\endgroup
\begin{picture}(250,76)(75,710)
\thicklines
\put( 80,740){\circle*{10}}
\put(120,740){\circle*{10}}
\put(160,740){\circle*{10}}
\put(200,740){\circle*{10}}
\put(240,740){\circle*{10}}
\put(280,740){\circle*{10}}
\put(320,740){\circle*{10}}
\put(240,780){\circle*{10}}
\put(240,740){\line( 0, 1){ 40}}
\put( 80,740){\line( 1, 0){240}}
\put( 80,710){\makebox(0,0)[b]{\smash{\SetFigFont{14}{16.8}{bf}1}}}
\put(120,710){\makebox(0,0)[b]{\smash{\SetFigFont{14}{16.8}{bf}2}}}
\put(160,710){\makebox(0,0)[b]{\smash{\SetFigFont{14}{16.8}{bf}3}}}
\put(200,710){\makebox(0,0)[b]{\smash{\SetFigFont{14}{16.8}{bf}4}}}
\put(240,710){\makebox(0,0)[b]{\smash{\SetFigFont{14}{16.8}{bf}5}}}
\put(280,710){\makebox(0,0)[b]{\smash{\SetFigFont{14}{16.8}{bf}6}}}
\put(320,710){\makebox(0,0)[b]{\smash{\SetFigFont{14}{16.8}{bf}7}}}
\put(255,775){\makebox(0,0)[b]{\smash{\SetFigFont{14}{16.8}{bf}8}}}
\end{picture}
\]
Furthermore, define the functions
$K^{\mbox{\scriptsize E}_8}_{t,s}$,
$A^{\mbox{\scriptsize E}_8}_{t,s}$,
$a^{\mbox{\scriptsize E}_8}_{t,s}$
and $s$ by their FT
\begin{eqnarray}
\hat{K}^{\mbox{\scriptsize E}_8}_{t,s}(x)&=&
\delta_{t,s} + \hat{s}(x)
\left(C^{\mbox{\scriptsize E}_8}_{t,s} - 2\delta_{t,s}\right)
\nonumber \\
\hat{A}^{\mbox{\scriptsize E}_8}_{t,s}(x)&=&
\left[\hat{K}^{\mbox{\scriptsize E}_8}(x)\right]^{-1}_{t,s}
\nonumber \\
\hat{a}^{\mbox{\scriptsize E}_8}_{t,s}(x)&=&
\hat{s}(x)
\hat{A}^{\mbox{\scriptsize E}_8}_{t,s}(x)\\
\hat{s}(x) &=& \frac{1}{2\cosh x} \nonumber
\end{eqnarray}
With these definitions, and after eliminating $\rho_0$,
 the integral equations (\ref{TBAE}) and
the energy expression (\ref{Energy}) take the form
\begin{eqnarray}
a_{1,t}^{\mbox{\scriptsize E}_8}(\alpha)
&=& \tilde{\rho}_t(\alpha)
+ \sum_{s=1}^8 A^{\mbox{\scriptsize E}_8}_{t,s} \ast \rho_s
\, (\alpha) \qquad t=1,\ldots,8, \nonumber \\
\frac{E}{N} &=&-\epsilon\sum_{t=1}^8
\int_{-\tau r/\pi}^{\tau r/\pi}
a_{1,t}^{\mbox{\scriptsize E}_8}(\alpha) \, \rho_t(\alpha)
\: d\alpha +\mbox{const}.
\label{BAE8}
\end{eqnarray}

We can now use (\ref{BAE8}) to study the scaling limit of the
model. In fact, all relevant calculations have already been
carried out in ref~\cite{Bazhanov90d} and we only need
to refer to the appropriate results therein.
To make the correspondence with ref~\cite{Bazhanov90d}
somewhat more transparent, let us give the
expression for the equilibrium free energy $F(T)$ of
the one-dimensional spin chain
at finite temperature $T$,
as it follows from (\ref{BAE8}) via standard TBA calculations
\cite{Yang-Yang},
\begin{equation}
\frac{F(T)}{N} = -\sum_{t=1}^8
\int_{-\tau r/\pi}^{\tau r/\pi}
a_{1,t}^{\mbox{\scriptsize E}_8}(\alpha) \,
T \log \left(1+\e^{-\beta \epsilon_t(\alpha)}\right)
\, d\alpha +\mbox{const},
\label{FE}
\end{equation}
where $\beta=1/T$ is the inverse temperature.
The functions $\epsilon_t=T\log(\tilde{\rho}_t/\rho_t)$ are
the solutions of the integral equation
\begin{equation}
\epsilon \delta_{1,t} s(\alpha) =
T \log \left(1+\e^{-\beta \epsilon_t(\alpha)}\right)
- \sum_{s=1}^8 K_{t,s}^{\mbox{\scriptsize E}_8}
\ast
T \log \left(1+\e^{\beta \epsilon_s(\alpha)}\right)
(\alpha).
\label{NLIE}
\end{equation}

The above two equations are equivalent to
(3$\cdot$20) and (3$\cdot$21) of ref~\cite{Bazhanov90d},
respectively, with their $\cal G$=E$_8$, $r=32$, $g=30$, $p=\ell = 1$,
their nome $q$ replaced by $q^{1/2}$ and
with their $\epsilon_j^a$ negated.
This last difference reflects the fact that our TBA equations are dual
to those of ref~\cite{Bazhanov90d} in the sence that the densities of
strings and holes are interchanged. From (\ref{FE}) and (\ref{NLIE})
it follows that for $T=0$
\begin{equation}
\begin{array}{lll}
i) & \epsilon =-1 \quad & \epsilon_t(\alpha) =
a_{1,t}^{\mbox{\scriptsize E}_8} \\
& & \\
ii) & \epsilon =+1 & \epsilon_t(\alpha) =
-\delta_{t,1} s(\alpha).
\end{array}
\end{equation}
The functions $|\epsilon_t(\alpha)|$ are the energies
of the excitations over the ground state.

For $\epsilon=-1$ the ground state is formed by type 0
strings. As was remarked after equation (\ref{zero}),
these strings have no holes for any state. Therefore the
Dirac sea is ``frozen'', and the excitations correspond to
the remaining eight string types. The phenomenon of ``freezing''
of the Dirac sea which can be interpreted as the confinement of
``holes'' has been first observed in
the TBAE calculations of ref~\cite{Bazhanov}
for the RSOS models of
Andrews, Baxter and Forrester~\cite{A-B-F}.

For $\epsilon=1$ the Dirac sea is formed by the type 1
strings, and the only excitations correspond to holes
in the Dirac sea. These excitations are of the kink type.

Now we consider the scaling limit. We introduce a dimensional
spacing parameter $d$ for our chain and
take the limit $N\to\infty$, $d\to 0$,
keeping the (dimensional) length of the chain $L=N d $ to
be macroscopically bigger than the correlation length:
$L>>R_c= q^{-\xi}d$, where $\xi$ is the index of the correlation
length.
In the scaling limit we thus have $d\sim q^{\xi}$,
$N>>q^{-\xi}$, $q\to 0$, and we
obtain the massive relativistic spectrum
of excitations.
To find this, one has to compute the energy dispersion law for the
physical excitations in the $q\to 0$ limit keeping the
rapidities $\alpha$ of the order of $\alpha_0=\tau r/\pi$, where
the functions $|\epsilon_t(\alpha)|$ have their minima.
Taking into account the correspondence in notation
discussed after (\ref{NLIE}), one gets from
(4$\cdot$1) and (4$\cdot$2) of ref~\cite{Bazhanov90d}
\begin{eqnarray}
i) & & \epsilon_t
\left(\case{30}{\pi} \beta + \alpha_0 \right) = m_t \cosh \beta +
o(q^{\xi}
) \nonumber \\
& & m_t = \mbox{const } S_t q^{\xi}, \quad \xi = \case{8}{15}
\nonumber \\
ii) & & \left|\epsilon_1
\left(\case{2}{\pi} \beta + \alpha_0 \right)
\right| = m \cosh \beta +
o(q^{\xi}) \label{Mass} \\
& & m = \mbox{const } q^{\xi}, \quad \xi = 8,
\nonumber
\end{eqnarray}
where $\beta$ here denotes the rapidity variable and
$S_t$ was defined just before equation (\ref{massratio}).

Using the scaling relation
$\xi = (2-2\Delta)^{-1}$ it is
seen that the values of $\xi$ in (\ref{Mass})
lead exactly to the dimensions of the leading perturbations
as given in (\ref{canddelta}).

The values of the central charges of the corresponding
(ultraviolet) conformal field theories listed in
(\ref{canddelta}) have also been previously calculated.
For regime $i)$ in \cite{Bazhanov90d,K-M,AlZamolodchikov}
and for regime $ii)$ in \cite{Bazhanov90d}.

Finally, the
$S$-matrix for regime {\it i)}, where all the string excitations
for $t=1\ldots 8$ correspond to distinct particles, can be found
straightforwardly from equation (\ref{BAE8}).
The result is \cite{Bazhanov90d}
\begin{equation}
S_{t,s}(\beta) = \exp\left\{ \ib
\int_0^{\infty} A_{t,s}^{\mbox{\scriptsize E}_8}(x) \,
\frac{\sin (30\beta x/{\pi})}{x} \: dx \right\},
\end{equation}
which coincides with Zamolodichkov's E$_8$ $S$-matrix
\cite{Zamolodchikov-A89a}.

For regime {\it ii)} the kink-kink $S$-matrix is of
the RSOS type related to the E$_7$ Lie algebra
\cite{Bazhanov90d}, and will be discussed elsewhere
\cite{BNW}.


\nsection{Summary and Conclusion}
In this paper we have established the final link between
Zamolodchikov's E$_8$ $S$-matrix of the critical Ising model
in a field \cite{Zamolodchikov-A89a}
and its underlying lattice model.
By making a conjecture for the possible string
solutions of its Bethe Ansatz equations,
we have derived a system of thermodynamic BAE for the
dilute A$_3$ lattice model of Warnaar et al. \cite{WNS}.
After a suitable transformation
we have recast these TBAE in terms of the root system
of the Lie algebra E$_8$.
These E$_8$ TBAE are found to be precisely those conjectured
earlier by Bazhanov and Reshetikhin \cite{Bazhanov90d}, and
using their results, the correspondence
between the dilute A$_3$ model and the E$_8$ $S$-matrix
is made.

To conclude
we mention that two more remarkable integrable $\phi_{1,2}$ perturbations
of CFT's are known, notably those related to $S$-matrices with
hidden E$_7$ (c=7/10) and E$_6$ (c=6/7) structure \cite{F-Z}.
Like the E$_8$ case, the underlying lattice models
of these integrable QFT's correspond to models in the
dilute A hierarchy.
The working for these two extra cases, corresponding
to dilute A$_4$ and A$_6$, respectively, as well as some
additional results for the dilute A$_3$ model will be the
subject of a future publication \cite{BNW}

\section*{Acknowledgements}
We wish to thank
M.~T.~Batchelor, R.~J.~Baxter, U.~Grimm, P.~A.~Pearce,
and N.~Yu.~Reshetikhin for interesting discussions.
We thank U.~Grimm, P.~A.~Pearce and Y.~K.~Zhou
for sending us their work prior to publication.
One the authors (VVB) thanks
the University of Amsterdam for hospitality during his
visit in the summer of 1992, when this work
has been initiated.
This work has been supported by the Stichting voor Fundamenteel
Onderzoek der Materie (FOM).


\end{document}